\documentclass[conference]{IEEEtran}
\IEEEoverridecommandlockouts
\usepackage{amsmath,graphicx}
\usepackage[colorlinks]{hyperref}
\usepackage{cite}
\usepackage{amsmath,amssymb,amsfonts}
\usepackage{algorithmic}
\usepackage{graphicx}
\usepackage{textcomp}
\usepackage{xcolor}

\usepackage{booktabs}

\usepackage{soul}

\usepackage{multirow}
\usepackage{adjustbox}
\usepackage{array}
\usepackage[utf8]{inputenc}

\def\ninept{\def\baselinestretch{1.025}\let\normalsize\small\normalsize}
\ninept

\def\BibTeX{{\rm B\kern-.05em{\sc i\kern-.025em b}\kern-.08em
    T\kern-.1667em\lower.7ex\hbox{E}\kern-.125emX}}
\begin{document}

% \title{\Huge Multi-Source Music Generation with Latent Diffusion}
\title{\fontsize{23.5}{26}\selectfont Multi-Source Music Generation with Latent Diffusion}

 % Debottam Dutta, Yu-lin Wei, Romit Roy Choudhury

\author{
    \IEEEauthorblockN{Zhongweiyang Xu,
                      Debottam Dutta,
                      Yu-Lin Wei,
                      Romit Roy Choudhury}
    \IEEEauthorblockA{Department of Electrical and Computer Engineering, University of Illinois Urbana-Champaign}
    % Optionally add emails here if needed
    % Emails: {zhongweiyang, ddutta, yulin, rrc}@illinois.edu
}

\maketitle

\begin{abstract}
 Most music generation models directly generate a single music mixture. To allow for more flexible and controllable generation, the Multi-Source Diffusion Model (MSDM) has been proposed to model music as a mixture of multiple instrumental sources (e.g. piano, drums, bass, and guitar). Its goal is to use one single diffusion model to generate mutually-coherent music sources, that are then mixed to form the music. Despite its capabilities, MSDM is unable to generate music with rich melodies and often generates empty sounds. Its waveform diffusion approach also introduces significant Gaussian noise artifacts that compromise audio quality. In response, we introduce a Multi-Source Latent Diffusion Model (MSLDM) that employs Variational Autoencoders (VAEs) to encode each instrumental source into a distinct latent representation. By training a VAE on all music sources, we efficiently capture each source's unique characteristics in a ``source latent''. The source latents are concatenated and our diffusion model learns this joint latent space. This approach significantly enhances the total and partial generation of music by leveraging the VAE’s latent compression and noise-robustness. The compressed source latent also facilitates more efficient generation. Subjective listening tests and Fréchet Audio Distance (FAD) scores confirm that our model outperforms MSDM, showcasing its practical and enhanced applicability in music generation systems. We also emphasize that modeling sources is more effective than direct music mixture modeling. Codes and models are available at \href{https://github.com/XZWY/MSLDM}{https://github.com/XZWY/MSLDM}. Demos are available at \href{https://xzwy.github.io/MSLDMDemo/}{https://xzwy.github.io/MSLDMDemo/}.
\end{abstract}

% \vspace{-1pt}
\begin{IEEEkeywords}
% \vspace{-4pt}
Music Generation, Latent Diffusion
\end{IEEEkeywords}

% \vspace{-3pt}
\section{Introduction}
% \vspace{-1pt}
\label{sec:intro}

 % Music Generation
Generative models show impressive performance not only in language and image modeling~\cite{gpt, dalle, gpt2}, but also show promising results in music generation. Music generation models usually fall into two categories: 1) Auto-regressive models and 2) Diffusion models. 

For auto-regressive models, WaveNet~\cite{wavenet} directly models scalar-quantized waveform samples, which allows for generating small musical fragments. However, due to the sample-level auto-regression, WaveNet has low sampling efficiency. One way to improve efficiency is to encode waveform samples to a discrete latent representation (tokens) with a much lower time resolution. These tokenizers~\cite{jukebox, encodec, soundstream} are usually variations of VQ-VAE~\cite{vqvae}, but are usually trained with perceptual adversarial loss~\cite{encodec, soundstream, hifigan, dac}. Then, the auto-regressive transformer models the sequence of tokens achieving higher efficiency. Among these models, JukeBox~\cite{jukebox} allows music generation conditioned on lyrics. More recently, text-to-music generation has shown significant progress based on this framework~\cite{musiclm, musicgen}.

On the other hand, diffusion-based music generation models also hold great potential, where the diffusion model is learned on some intermediate representation. Noise2Music~\cite{noise2music} uses diffusion models to generate an intermediate representation, either downsampled waveforms, or Mel-Spectrogram features, and then decodes the intermediate representation to music waveform by a cascader or vocoder. Then, because of the success of latent diffusion in image generation~\cite{ldm}, music generation also follows this path. DiffSound~\cite{diffsound} first trains a spectrogram VQ-VAE tokenizer as the intermediate representation, and then uses a discrete diffusion model to model the token sequence. \cite{makeanaudio, audioldm1, audioldm2, musicldm} are using spectrogram-domain (variational) autoencoder's continuous latent as an intermediate representation for diffusion, while~\cite{stableaudio, jen1} uses waveform-domain VAE's latent as the diffusion target. Moûsai~\cite{mousai} proposes a spectrogram encoder learned by diffusion magnitude autoencoding (DMAE) and then trains another diffusion model on the encoder's latent. Further, with a waveform-domain VAE, StableAudio2~\cite{stableaudio2} achieves full-song generation by modeling the VAE latent with diffusion.

 % Multi-Source Music Generation (STEMGEN + MSDM)
Although rapid progress has been made in music generation, most models directly generate the whole music piece, which is a mixture of individual sources. However, the individual sources cannot be disentangled from the mixture. Ideally, a music generation method should be able to generate the individual music sources that together form a piece of music, similar to a human's music composition process. This will allow the music to be more interpretable and controllable (e.g., the piano can be made louder than the drums). To solve this problem, one class of methods directly models the musical notes or midi representations in a multi-track manner~\cite{symbolic_diffusion, musegan}. However, the generated notes or midi sequence need to be later decoded to a single waveform using synthesizers. The other type of research directly learns to model several music tracks directly. StemGen~\cite{stemgen} uses a masked language model on Encodec tokens to generate any single instrument source given a music context. \cite{bass} uses the latent diffusion model to generate bass companions conditioned on mixtures, while SingSong~\cite{singsong} generates background companions given the vocal source. Most recently, MSDM~\cite{msdm} has been proposed to simultaneously model four instrument sources (piano, drums, bass, guitar) with a single waveform-domain diffusion model, and GMSDI~\cite{gmsdm} has generalized MSDM by training on text-conditioned diffusion models allowing adaption to any music dataset using text descriptions. The closest work to ours is multi-track MusicLDM~\cite{tornike1, tornike2}, which is simultaneous to this paper, some implementation differences exist. 

\begin{figure*}[ht]
\vspace{-0.2cm}
    \centering
          \includegraphics[scale=0.8]{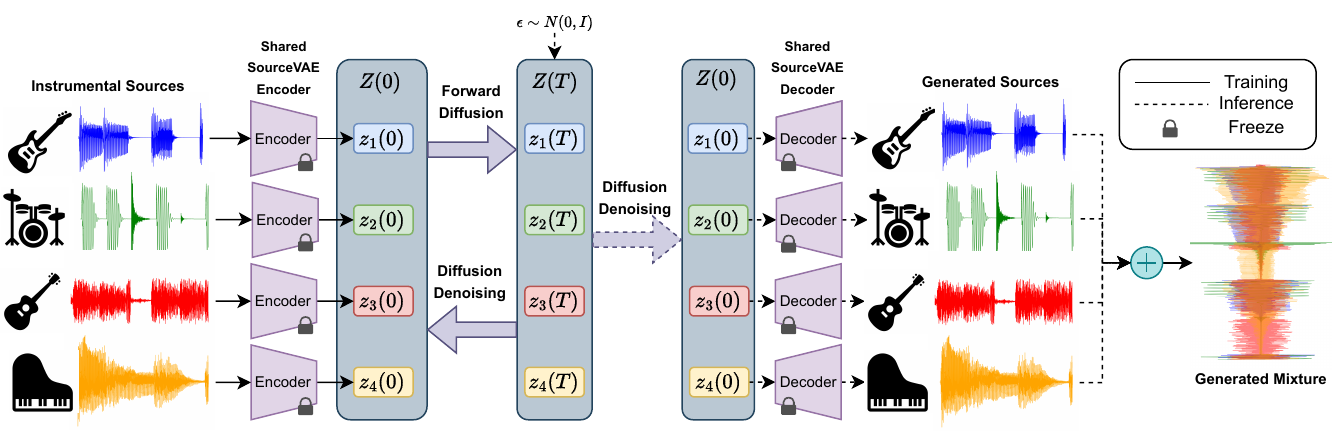}
          \vspace{-15pt}
    \caption{An Overview of the proposed MSLDM framework.}
    \label{fig:Overall}
\vspace{-12pt}

\end{figure*}

 % our model, contribution
In this paper, we propose to simultaneously model four different instrumental sources (piano, drums, bass, guitar) jointly, with a single multi-source latent diffusion model (MSLDM), as shown in~Fig.\ref{fig:Overall}. We first train a shared SourceVAE on all the sources to perceptually compress the source audio, and then use this VAE's encoder to extract the latent feature of each source. We then apply diffusion to model the generation of the latents of the sources. We claim that 1) with the VAE compressing the source audio into a compact latent, the diffusion can better model semantic and sequential information like melodies and the harmony between sources, and 2) modeling individual sources is better than direct modeling mixtures. Our result in both subjective human evaluation and objective FADs validates our claim.

% \vspace{-8pt}
\section{Models}
% \vspace{-7pt}

Fig.~\ref{fig:Overall} shows the training and inference pipeline of our model. Just like any latent diffusion model, our model involves two blocks: (1) SourceVAE, which is trained like a VAE but with an adversarial loss for perceptual compression. (2) A diffusion model simultaneously models all the sources' latents concatenated together as one latent. Inference includes two sub-tasks as well: (1) Total generation allows unconditional generation of all instrumental sources at the same time and (2) Partial generation allows the generation of companion sources given any combinations of instrumental sources (e.g., generate bass and guitar to accompany given piano and drums).

% \subsection{Training}
% \vspace{-8pt}
\subsection{SourceVAE}\label{sec:sourcevae}
The SourceVAE aims to compress waveform-domain instrumental sources into a compact latent space, while still ensuring perceptually indistinguishable reconstruction. This is usually achieved by adversarial training with carefully designed discriminators. We borrow this training framework and model architecture from the DAC~\cite{dac} neural audio codec. DAC is a state-of-the-art waveform-domain neural audio codec trained with both reconstruction and adversarial losses. It can encode, quantize, and decode audio with superior quality. However, we want our latent to be noise-robust and continuous, so we remove the vector quantization module, constrain the intermediate latent size, and add a small KL-divergence loss term as used in vanilla VAEs~\cite{VAE}. The KL-divergence loss is to ensure a noise-robust latent space. For the encoder and decoder, we use DAC's default 24kHz model but with a new latent size of $C=80$. Given an instrumental source $s\in\mathbb{R}^{N}$ with $N$ samples, the encoder encodes the waveform to a posterior $\Psi_{enc}(s)=\mathcal{N}(\cdot|\mu_z(s), \Sigma_z(s))$, where $\mu_z(s)\in\mathbb{R}^{C\times\frac{N}{D}}$ is the posterior mean of the latent and $\Sigma_s$ is the corresponding posterior covariance. $D$ is the time-domain downsampling factor of the encoder, which is $320$ in DAC. Then any $z\sim\mathcal{N}(\cdot | \mu_z(s), \Sigma_z(s))$, processed by the decoder $\psi_{\text{dec}}$, should reconstruct $s$ with good quality. To extract latent features, we take the posterior mean $z_s=\mu_z(s)$. During training, the loss is as shown below:
% \vspace{-3pt}
\begin{equation}
   L_{\text{SourceVAE}} = \lambda_1L_{\text{Mel}} + \lambda_2L_{\text{feature}} + \lambda_3L_{\text{adversarial}} + \lambda_4L_{\text{KL}}
    % \vspace{-7pt}
\end{equation}
 % \vspace{-2pt}
where $L_\text{Mel}$, $L_{\text{feature}}$, $L_{\text{adversarial}}$ are Mel-reconstruction loss, feature matching loss, and adversarial loss, respectively, as in the DAC training framework. Also $\lambda_1=15$, $\lambda_2=2$, $\lambda_3=1$ according to DAC. $L_{\text{KL}}$ is the KL-Divergence between the VAE encoded posterior and $\mathcal{N}(0, I)$ and we set $\lambda_4=10$. The implementation details are available in our source code.

\vspace{-2pt}
\subsection{Multi-Source Latent Diffusion}\label{sec:msldm}
\vspace{-1pt}
In our setup, assume any music piece $x\in\mathbb{R}^N$ is a mixture of $K$ instrumental sources: $x=\sum_{k=1}^{K}s_k$, where $S=(s_1, s_2, ..., s_K)\in\mathbb{R}^{K\times N}$ coherently added together to form the musical mixture $x$. Our goal is to sample from the distribution of $S$ to get multi-source music. Instead of directly modeling the generation of $S$ as in \cite{msdm}, we propose to model the generation of $Z_S=(z_{s_1}, z_{s_2}, ..., z_{s_K})\in\mathbb{R}^{K\times C\times\frac{N}{D}}$, where $z_{s_i}\in\mathbb{R}^{C\times\frac{N}{D}}$ is the SourceVAE's latent of $s_i\in\mathbb{R}^{N}.$ For the notation to be more abbreviated, we will ignore the subscript $s$ for the latent, so we are modeling the generation of $Z = (z_1, z_2, ..., z_K)\in\mathbb{R}^{K\times C\times\frac{N}{D}}$.

We model the generation of $Z = (z_1, z_2, ..., z_K)$ with a score-based diffusion model~\cite{score}. Following EDM~\cite{karras}, with the diffusion schedule $\sigma(t)=t$, the forward diffusion process is defined by:
\begin{equation}\label{eq:forward}
d{Z}(t) = -\sigma(t) \nabla_{{Z}(t)} \log p({Z}(t)) dt    
\end{equation}
% \begin{equation}\label{eq:forward}
% d{S}^\tau = -\sigma(\tau) \nabla_{{S}^\tau} \log p({S}^\tau) dt    
% \end{equation}
where $Z(t) = \mathcal{N}(Z(0), \sigma^2(t)I), Z(0)=Z$. Then with an ODE solver, we sample $Z$ by solving the backward process:
\begin{equation}\label{eq:backward}
d{Z}(t) = \sigma(t) \nabla_{{Z}(t)} \log p({Z}(t)) dt    
\end{equation}
% \begin{equation}\label{eq:backward}
% d{S}^\tau_{1:K} = \sigma(\tau) \nabla_{{S}^\tau_{1:K}} \log p({S}^\tau_{1:K}|X) dt    
% \end{equation}
  We approximate the score $\nabla_{{Z}(t)}\log p({Z}(t))$ with a neural network $S^\theta(Z(t), \sigma(t))$ and then train the score matching loss following the practice in EDM~\cite{karras}, i.e. $\sigma_{data}=0.4, p_{train}(\sigma)=\mathrm{Uniform}(0, 3)$.

\vspace{-3pt}
\subsection{Inference}\label{sec:inference}
\vspace{-1pt}
The inference pipeline is marked by the dashed objects in Fig.~\ref{fig:Overall}. During sampling, we use the same sampler for MSDM~\cite{msdm}. The sampler is an Euler method-based ODE solver to integrate Eq.~\ref{eq:backward} with some stochasticity controlled by the parameter $s_{\text{churn}}$, as proposed in EDM~\cite{karras}. We use $\sigma_{min}=0.01, \sigma_{max}=3, \rho=7, s_{churn}=20, n_{steps}=150$, also following the configuration in EDM. 

% \vspace{-1pt}
\subsubsection{Total Generation}
% \vspace{-1pt}

\label{sec:total}
The total generation inference is straightforward. Starting from randomly sampled white noise $Z(T)\sim\mathcal{N}(\cdot | 0, \sigma_{max}^2I)$, the diffusion sampling process gradually transforms $Z(T)$ to $Z(0)$. Then instrumental source latent $z_1, z_2, ..., z_K$ are extracted from $Z(0)$, and are further decoded independently by the SourceVAE decoder to get the generated source waveforms $\{s_i\in\mathbb{R}^{N}| s_i = \psi_{dec}(z_i), i\in[1,2,..., K]\}$. These generated sources could then be added to form a mixture of music pieces $x = \sum_{k=1}^{K}s_i$.

% \vspace{-8pt}
\subsubsection{Partial Generation}
% \vspace{-1pt}
\label{sec:partial}
Partial generation is the task of generating complementary sources given some existing ones to condition on. Assume a subset of instruments are given by the indices $I \subset \{1,2,...K\}$, and the corresponding given sources are denoted by $S_I = \{s_i\}_{i\in I}$. Then the complementary sources to generate are indexed by $\bar{I} = \{1,...,K\}\backslash I$, which means the source to generate are $S_{\bar{I}} = \{s_i\}_{i\in \bar{I}}$. Since our diffusion model works in the latent domain, we first use SourceVAE to encode each source in $S_I$ to latent $Z_I = \{z_i|z_i = \mu_z(s_i), i\in I\}$. The task is to generate $Z_{\bar{I}}$ conditioned on $Z_{I}$, so we need the conditional score $\nabla_{{Z}_{\bar{I}}(t)} \log p(Z_{\bar{I}}(t)|Z_{I}(t))$ for sampling. Following diffusion-based imputation~\cite{score} and MSDM~\cite{msdm}, we could estimate the condition score by:
% \begin{align}\label{eq7}
% \nabla_{{Z}_{\bar{I}}(t)} \log p(Z_{\bar{I}}(t)|Z_{I}(t)) &\approx \nabla_{{Z}_{\bar{I}}(t)} \log p([Z_{\bar{I}}(t), \hat{Z}_{I}(t)]) \\
% &\approx S^{\theta}([Z_{\bar{I}}(t), \hat{Z}_{I}(t)], \sigma(t))
% \end{align}

\vspace{-10pt}
{\small % This makes the text within the braces smaller
\begin{align}
\nabla_{{Z}_{\bar{I}}(t)} \log p(Z_{\bar{I}}(t)|Z_{I}(t)) &\approx \nabla_{{Z}_{\bar{I}}(t)} \log p([Z_{\bar{I}}(t), \hat{Z}_{I}(t)]) \\
&\approx S^{\theta}([Z_{\bar{I}}(t), \hat{Z}_{I}(t)], \sigma(t))\label{eq:score}
\end{align}
}

where $\hat{Z}_{I}(t)$ is sampled from $\mathcal{N}(\cdot|{Z}_{I}(0), \sigma^2(t))$. Then using Eq.~\ref{eq:score}, we can use the sampling method in Sec.~\ref{sec:inference} to solve the following ODE (similar to Eq.\ref{eq:backward}) initialized from $Z_{\bar{I}}(T)\sim\mathcal{N}(\cdot | 0, \sigma_{max}^2I)$:
\begin{equation}
    d{Z_{\bar{I}}}(t) = \sigma(t) S^{\theta}([Z_{\bar{I}}(t), \hat{Z}_{I}(t)], \sigma(t)) dt    
\end{equation}
With ${Z_{\bar{I}}}(0)$ sampled, the partially generated sources $S_{\bar{I}}$ can be decoded by the SourceVAE decoder: $S_{\bar{I}} = \{s_i\in\mathbb{R}^{N}| s_i = \psi_{dec}(z_i), z_i\in Z_{\bar{I}}\}$. Then all the conditional sources and partially generated sources are added to form the final music piece $x$.

% $\nabla_{{Z}_{\bar{I}}(t)} \log p(Z_{\bar{I}}(t)|Z_{I}(t))$ with $\nabla_{{Z}_{\bar{I}}(t)} \log p([Z_{\bar{I}}(t), \hat{Z}_{I}(t)])$, where $\hat{Z}_{I}(t)$ is sampled from $\mathcal{N}(\cdot|{Z}_{I}(t), \sigma(0))$

% which could be estimated by our network $S^{\theta}([Z_{\bar{I}}(t), \hat{Z}_{I}(t)], \sigma(t))$
% \vspace{-3pt}
\section{Experiments and Dataset}

% \vspace{-3pt}
\subsection{Dataset}
% \vspace{-1pt}

We use the same dataset as MSDM~\cite{msdm}, namely the \texttt{slakh2100} music dataset~\cite{slakh}. slakh2100 is a MIDI synthesized music dataset with 145 hours of music, containing both the mixed music and the individual tracks labeled with instrument class. Same as MSDM, we use $K=4$ main tracks which are piano, drums, bass, and guitar for multi-source modeling. For fair comparison, we use the identical sampling rate of 22,050Hz.

% \vspace{-8pt}
\subsection{SourceVAE}
% \vspace{-1pt}

The SourceVAE mentioned in Sec.~\ref{sec:sourcevae} is a 1D-CNN-based encoder-decoder architecture coupled with a DAC loss and a KL-divergence loss. The encoder and decoder all follow the final setup of the DAC architecture for 24kHz. The intermediate latent dimension for SourceVAE is set to be $C=80$ and the encoder has a downsampling rate of $D=320$ for the temporal dimension. For training, we use a batch size of 28 and train on one-second-long single-instrumental segments for 100k steps. All other SourceVAE training configurations are the same as the DAC paper with code available at \href{https://github.com/descriptinc/descript-audio-codec}{https://github.com/descriptinc/descript-audio-codec}. 

% \vspace{-8pt}
\subsection{Latent Diffusion and Unet Architecture}
% \vspace{-1pt}

As mentioned in Sec.~\ref{sec:msldm}, the score estimation network $S^\theta(Z(t), \sigma(t))$ learns a function mapping as shown below:
\begin{equation}
    S^{\theta}(Z(t), \sigma(t)) : \mathbb{R}^{K \times C \times \frac{N}{D}} \times \mathbb{R} \rightarrow \mathbb{R}^{K \times C \times \frac{N}{D}}
\end{equation}
To accommodate the 1D-Unet architecture used in MSDM, we concatenate the K source latent channel-wise, so the new channel dimension becomes $KC$, and the input to the 1D-Unet is $Z'(t)\in \mathbb{R}^{KC \times \frac{N}{D}}$, which is a reshaped version of $Z(t)$. $KC$ is treated as the channel dimension and $\frac{N}{D}$ is treated as the temporal dimension for the Unet. 
In our setup, $K=4, C=80, D=320$. We train our diffusion model on segments with $N=327672$ samples (about 15 seconds). For the 1D-Unet architecture, similar to MSDM, we adapt the architecture used in Moûsai~\cite{mousai} but with some modifications. We set the input channel dimension to be $KC=320$ and then we experiment on two different configurations. We call one model MSLDM and one larger model MSLDM-\texttt{Large}. The MSLDM contains 6 nested U-Net blocks with increasing channels [1024, 2048, 4096, 4096, 4096, 4096]. The downsampling factor for the blocks is [1,1,2,1,1,2]. The self-attention blocks are used for all the blocks except the first one. 12 Attention heads are used, and each head is 64 dimensional. For MSDLM-Large, the Unet contains 8 layers where the corresponding output channel dimensions are [1024, 2048, 4096, 4096, 4096, 4096, 4096, 4096]. The corresponding downsampling factors are [1,1,2,1,1,2,2,2]. All blocks contain self-attention blocks (except the first one) and each attention block contains 12 attention heads that are each 128 dimensional. The Unet and diffusion code setup are adapted from \href{https://github.com/archinetai/audio-diffusion-pytorch/tree/main}{audio-diffusion-pytorch}. Similar to MSDM, we train the diffusion model with a batch size 16 and a learning rate of 2e-5 for 400k steps.

% For the MixtureVAE trained for MixLDM, the training is exact the same as SourceVAE except it's trained on the mixture segments instead of single-instrumental segments.

All the model training is performed on a single RTX A6000 GPU with 48GB of VRAM. Further details can be found in our \href{https://github.com/XZWY/MSLDM}{code}.
% \vspace{-1pt}
\section{Evaluation Metrics and Results}
% \vspace{-1pt}

We evaluate two tasks: (1) Total generation and (2) Partial generation as mentioned in Sec.~\ref{sec:inference}. For both tasks, we evaluate the FAD score~\cite{fad} and human subjective score with a listening test. We compare our performance against 3 baselines.
% \vspace{-8pt}
\subsection{Baseline Models}
% \vspace{-1pt}

The first baseline is the \textbf{MSDM}~\cite{msdm} model. To show the effectiveness of modeling sources instead of mixtures, we design another baseline called \textbf{MixLDM}, where the latent diffusion model directly models the latent of music mixture, instead of instrumental sources. This is a more common practice in diffusion-based audio/music generation, where the model directly models the music mixture. For training this model, we first train a MixtureVAE which is the same as SourceVAE except that it is trained on Mixture music. The latent size $C=320$, so that the diffusion model's input is of the same dimension as MSLDM. To claim that our model generates sources that are mutually coherent (i.e. in harmony with each other), we design one baseline called \textbf{ISLDM} (Independent Source Latent Diffusion Model), where we train four independent diffusion models on four instruments' latents, respectively, so each model can generate one single instrument. For each single source model, the input channel size becomes $C=80$, and all the other training parameters are the same as MSLDM. Since all the single source models are independent of each other, they cannot generate mutually coherent sources. This is set as a baseline to validate our model's abilities to generate mutually coherent sources. All models' parameters and inference time to generate one 12-second mixture on a single RTX A6000 GPU are listed in Table~\ref{tab:models}.
\begin{table}[h]
    \centering
    \begin{tabular}{lcc}
        \toprule
        Model & \# parameters (M) & Inference Time (S) \\
        \midrule
        MSDM & 405 & 7.92 \\
        MixLDM & 364 & 5.44 \\
        ISLDM & 364$\times$4 & 5.44$\times$4 \\
        MSLDM-Ours & 364 & 5.44 \\
        MSLDM-Large-Ours & 1654 & 7.44 \\
        \bottomrule
    \end{tabular}
    \vspace{2pt}
    \caption{Model parameters and inference time for generating a 12-second music mixture.}
    \label{tab:models}
    \vspace{-25pt}
\end{table}

\begin{table*}[h]
\centering
% \vspace{-0.1cm}
\caption{\textbf{sub-FAD for Partial Generation.} The \texttt{sub-FAD} (lower is better) is reported for any source combinations (B: Bass, D: Drums, G: Guitar, P: Piano), where \textbf{BD} means conditioned on piano and guitar, the task is to generate Bass and Drums.}
% \vspace{5pt}
\label{tab:sub-fad-partial}
\renewcommand{\arraystretch}{1.2} % Increase row height
\begin{tabular}{lccccccccccccccc}
\toprule
\textbf{Model} & \textbf{B} & \textbf{D} & \textbf{G} & \textbf{P} & \textbf{BD} & \textbf{BG} & \textbf{BP} & \textbf{DG} & \textbf{DP} & \textbf{GP} & \textbf{BDG} & \textbf{BDP} & \textbf{BGP} & \textbf{DGP} & \textbf{Overall} \\
\midrule
MSDM               & 0.23 & 0.75 & \textbf{0.18} & 0.49 & 1.75 & 0.75 & 1.40 & 1.30 & 1.40 & 1.77 & 3.13 & 2.92 & 5.54 & 3.51 & 1.79 \\
ISLDM              & 0.30 & 1.41 & 0.75 & 0.42 & 1.52 & 1.14 & 0.76 & 1.56 & 1.76 & 1.33 & 1.85 & 2.03 & 1.78 & 2.17 & 1.34 \\
MSLDM       & 0.24 & 1.27 & 0.38 & \textbf{0.32} & 1.22 & 0.81 & 0.64 & 1.00 & 0.92 & \textbf{0.98} & 1.44 & 1.57 & 1.48 & 1.43 & 0.98 \\
MSLDM-Large & \textbf{0.14} & \textbf{0.51} & 0.23 & 0.41 & \textbf{0.56} & \textbf{0.49} & \textbf{0.61} & \textbf{0.59} & \textbf{0.66} & 1.05 & \textbf{0.81} & \textbf{1.01} & \textbf{1.40} & \textbf{1.25} & \textbf{0.70} \\
\bottomrule
\end{tabular}
\vspace{-10pt}
\end{table*}

\begin{figure}[ht]
% [H]
  \centering
  \includegraphics[width=\linewidth]{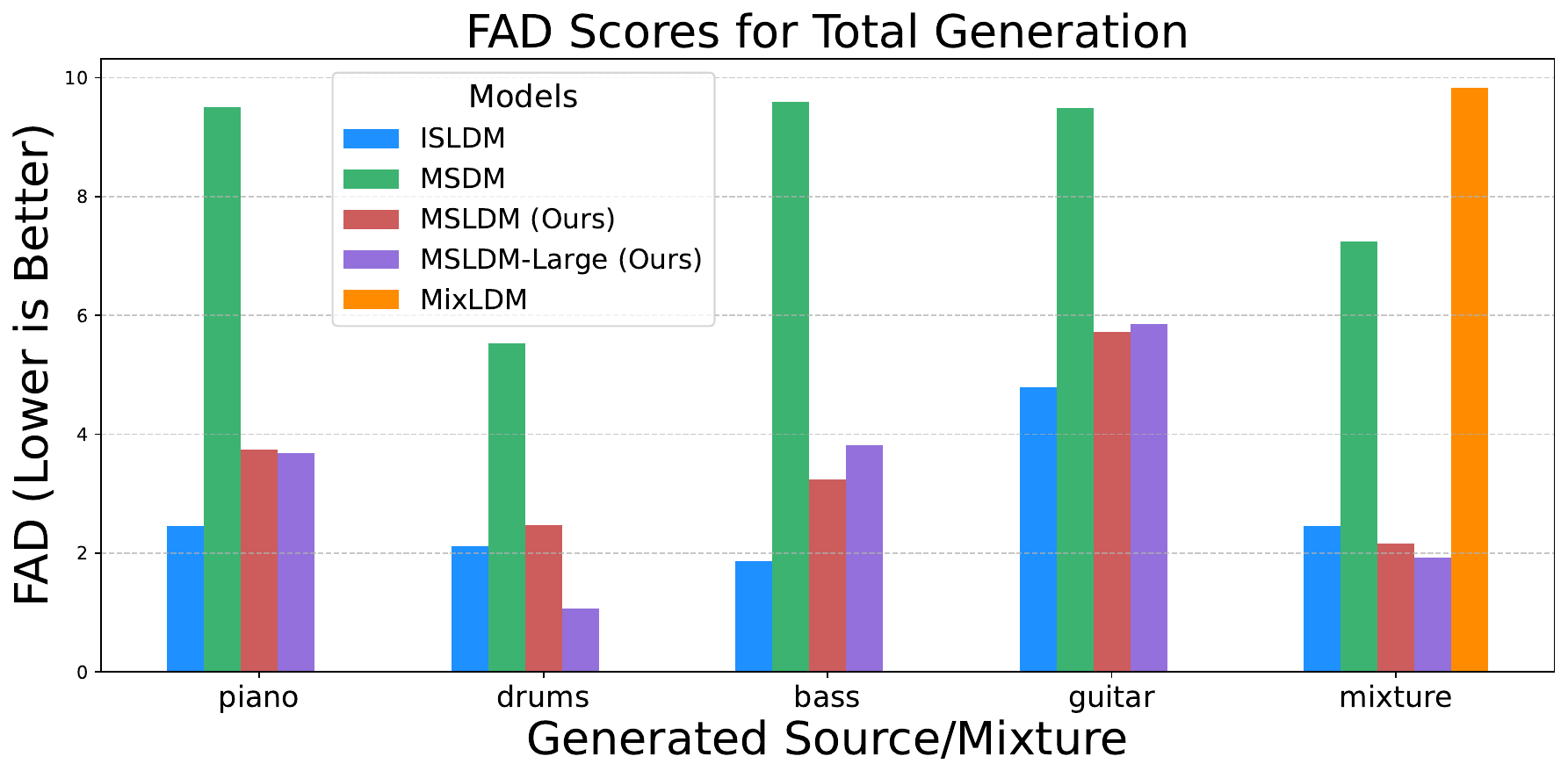}
\vspace{-20pt}
  \caption{Total Generation FAD.}
  \label{fig:fad_totl}
  \vspace{-10pt}
\end{figure}
  % \vspace{-20pt}

% \vspace{-5pt}
\subsection{Fréchet Audio Distance} (FAD)
% \vspace{-5pt}
We use the Fréchet Audio Distance (FAD)~\cite{fad} with VGGish feature~\cite{vggish} as the objective metric to evaluate total generation and partial generation. We use 200 chunks of music segment for the FAD calculation, where each segment is about 12 seconds, as in MSDM~\cite{msdm}. We randomly sample 200 12-second long music segments from the slakh2100 test set as the reference set.

\begin{figure}[ht]

% [H]
  \centering
  \includegraphics[width=\linewidth]{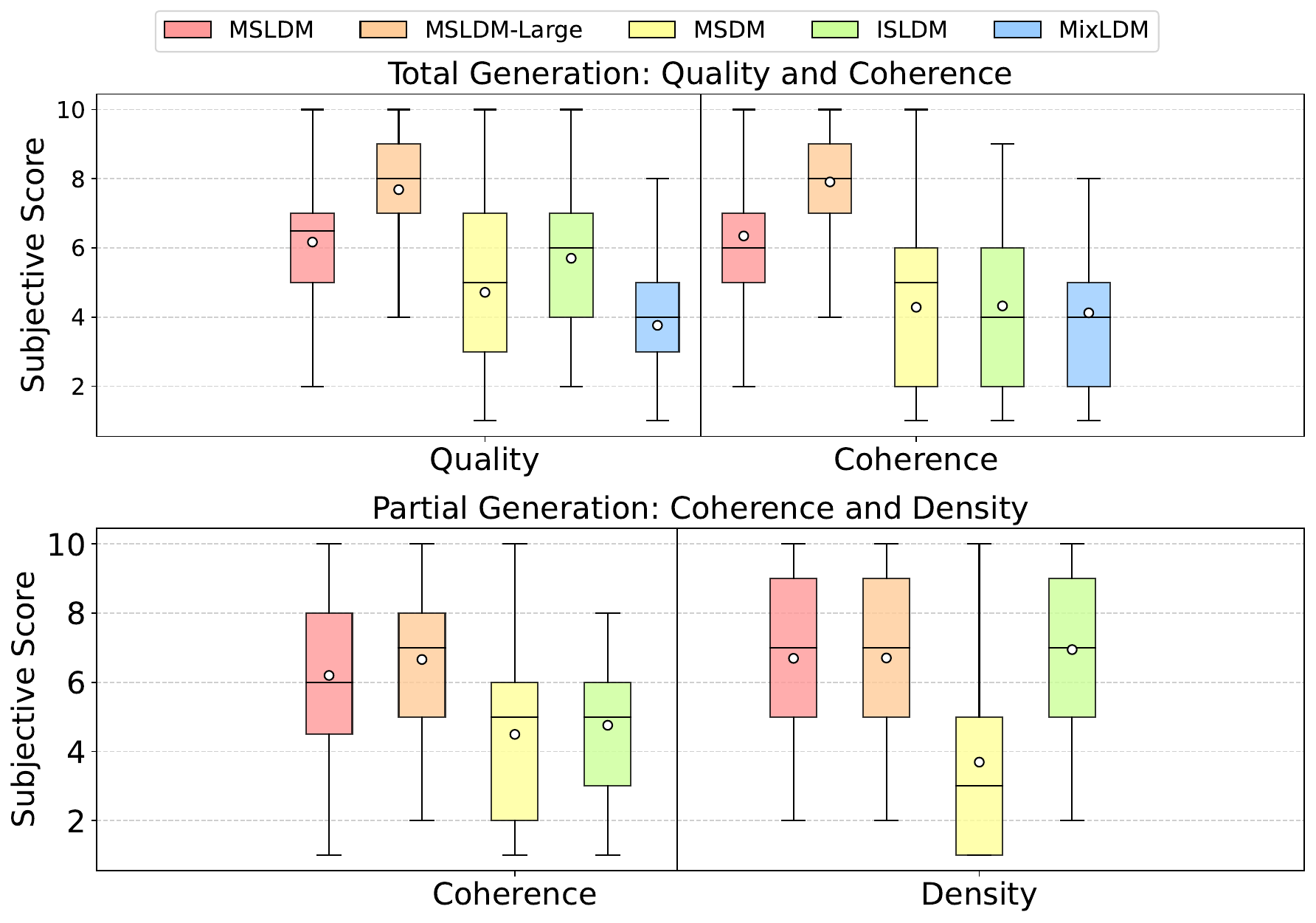}
\vspace{-15pt}
  \caption{Subjective listening test result.}
  \vspace{-10pt}
  \label{fig:subjective}
\end{figure}
% \vspace{-20pt}

For the task of \textbf{total generation}, we evaluate the generation of all the instrumental sources (piano, drums, bass, guitar) and the summed mixture (note: MixLDM can only be reported for the mixture). The FAD results are reported in Fig.~\ref{fig:fad_totl}. First, for single-source generation, ISLDM in general has the lowest FAD, showing the best generation performance. This is reasonable because each independent diffusion model only needs to model one single source. Then our proposed MSLDM and MSLDM-Large also show promising results, beating the MSDM baseline by a large margin. The drums generated by MSLDM-Large even achieve lower FAD than ISLDM. For mixture generation, MixLDM exhibits the worst performance. This confirms our belief that modeling the mixture directly is much harder than modeling sources. Observe that ISLDM, MSLDM, MSLDM-Large all outperform MSDM, demonstrating the efficacy of latent diffusion in modeling the harmony in realistic music mixtures. In fact, MSLDM and MSLDM-Large demonstrate lower FAD than ISLDM, showing that our model successfully generates sources in harmony. Note that when generating each single source, MSLDM performs worse than ISLDM for all instruments, but when sources are added together to form mixtures, MSLDM has lower FAD, showing that it is able to model the inter-source harmony.

For \textbf{partial generation}, we use the \texttt{sub-FAD} as the metric, similar to MSDM. 
\texttt{sub-FAD} calculates the FAD on a reference set and an evaluation set, where the reference set is the real music mixture, and the evaluated set is the mixture formed by mixing partially generated samples and originally given samples. For the dataset used for \texttt{sub-FAD} calculation, we again sample 200 12-second music segments from the slakh2100 test set, but we make sure that all the segments contain four instruments. The \texttt{sub-FAD} result is shown in Table~\ref{tab:sub-fad-partial}, where the results are shown for all models and all partial generation setups. The ISLDM model in this case is only generating the target sources using source-independent models. Overall, MSLDM-Large and MSLDM show much better performance than ISLDM and MSDM, with \texttt{sub-FADs} smaller than 1. Interestingly, the ISLDM's overall \texttt{sub-FAD} is lower than MSDM, implying that even though ISLDM cannot generate mutually coherent sources based on given sources, it is able to model single-source melody much better than MSDM. Across all the detailed setups, we see that MSLDM is consistently better than MSDM and ISLDM, except for the guitar.

% \vspace{-3pt}
\subsection{Subjective Listening Test}
% \vspace{-1pt}
\label{sec:results}
% \vspace{-1pt}
To complement the objective metrics, we also design subjective listening tests with human evaluation. We follow the exact test design in MSDM~\cite{msdm} for total and partial generation.

\textbf{Total Generation}: Each model generates 10 segments of 12-second music mixture samples. Then for each sample in the test, the participant is asked to rate the `quality' and `coherence', with a score from 1 to 10 (higher the better). `Quality' corresponds to how realistic the music is (considering white noise is the least realistic) and `coherence' corresponds to how mutually coherent the sources inside the mixture are. 
The evaluation scores are shown in the upper half of Figure~\ref{fig:subjective}. MSLDM and MSLDM-Large lead other baselines by a large margin in both quality and coherence. Also, MSLDM-Large exhibits a higher score than MSLDM, showing that large model size results in better performance. ISLDM's quality is better than MSDM, but its coherence is the worst because all the sources are generated independently. MixLDM has the worst quality and coherence, showing the difficulty in directly modeling the music mixture. In general, users find that our MSLDM model is consistently better than MSLDM, ISDM, MixLDM in both quality and coherence.

\textbf{Partial Generation}: We randomly sample 10 samples in the test set. Then, for each sample, the target source types (e.g., want to generate piano and drums given the two other instruments) are randomly sampled. Finally, the models are used to sample the specified source type for each data sample. We give the participants three music segments (the music to condition on, the partially generated music, and the synthesized mixture), and then ask them to rate 1-10 for `coherence' and `density'. The `density' corresponds to how dense (or sparse) the generated instruments are, in the 12-second segment (10 means all target instruments are generated for the whole segment). 
The results are shown in the lower half of Fig.~\ref{fig:subjective}. 
We observe that MSLDM and MSLDM-Large produce the highest coherence score, achieving the best performance in generating mutually-coherent sources that match the given ones. Again, ISLDM shows the worst coherence because sources are independent. For density, MSLDM, MSLDM-Large, and ISLDM all have high scores, showing the ability to generate rich companions. However, MSDM shows the worst performance in density because it often generates small and naive music segments, or even empty sources at times.

% The subjective scores are shown in Figure~\ref{fig:subjective}. For total generation, MSLDM and MSLDM-Large lead other baselines by a large margin in both quality and coherence. Also, MSLDM-Large exhibits a higher score than MSLDM, showing that large model size results in better performance. ISLDM's quality is better than MSDM, but its coherence is the worst because all the sources are generated independently. MixLDM has the worst quality and coherence, showing the difficulty in directly modeling the music mixture. In general, users find that our MSLDM model is consistently better than MSLDM, ISDM, MixLDM in both quality and coherence.

% For partial generation, we can see that MSLDM and MSLDM-Large produce the highest quality, showing the best performance in generating consistent sources that match the given ones. Again, ISLDM shows the worst quality because sources are independent. For density, MSLDM, MSLDM-Large, and ISLDM all have high scores, showing the ability to generate rich companions. However, MSDM shows the worst performance in density because it often generates small and naive music segments, or even empty sources at times.

Overall, both the objective metrics and subjective listening test show the MSLDM outperforms MSDM, ISLDM, MixLDM in terms of generating multi-source consistent music pieces. Compared with MSDM, MSLDM is better in terms of generating denser, more melodic, and more harmonious music. Compared with MixLDM, we showed that modeling sources is much easier than directly modeling music mixtures. Compared with ISLDM, we showed that our model is able to model inter-source dependency or harmony between different musical instruments.

% \vspace{-5pt}
\section{Conclusion and Future Work}
% \vspace{-2pt}
In this paper, we propose a multi-source diffusion model to jointly model the generation of multiple instruments together. Both objective and subjective metrics show better results in the tasks of total and partial generation, implying MSLDM is able to efficiently model melody and inter-source relations. Future research needs to advance MSLDM for weakly-supervised music separation and generalize our model to more instruments.

\newpage
% \newpage
\bibliographystyle{IEEEbib}
\bibliography{main}

\end{document}